\documentclass[aps,preprint,preprintnumbers,amsmath,amssymb,floatfix,showpacs]{revtex4}
\usepackage{graphicx}
\usepackage{epsfig}

\usepackage{amsbsy,bm}

\newcommand{\bpsi}{\boldsymbol{\psi}}
\newcommand{\bs}{\boldsymbol}
\newcommand{\mx}{\mathbf}

\begin{document}

\title{Solution of the Schr\"odinger equation using exterior complex scaling and fast Fourier transform}

\author{Vladislav V. Serov and Tatiana A. Sergeeva}
\affiliation{
Department of Theoretical Physics, Saratov State University, 83
Astrakhanskaya, Saratov 410012, Russia}

\date{\today}

\begin{abstract}
The split-operator pseudo-spectral method based on the fast Fourier transform (SO-FFT) is a fast and accurate method for the numerical solution of the time-dependent Schr\"odinger-like equations (TDSE). As well as other grid-based approaches, SO-FFT encounters a problem of the unphysical reflection of the wave function from the grid boundaries. Exterior complex scaling (ECS) is an effective method widely applied for the suppression of the unphysical reflection. However, SO-FFT and ECS have not been used together heretofore because of the kinetic energy operator coordinate dependence that appears in ECS applying.
We propose an approach for the combining the ECS with SO-FFT for the purpose of the solution of TDSE with outgoing-wave boundary conditions. Also, we propose an effective ECS-friendly FFT-based preconditioner for the solution of the stationary Schr\"odinger equation by means of the preconditioned conjugate gradients method.
\end{abstract}



\maketitle

\section{Introduction}

A numerical approach based upon the split-operator method and the fast Fourier transform (SO-FFT) is widely used for the solution of equations of the form of time-dependent Schr\"odinger equation (TDSE). 
Such equations commonly appear in the consideration of the wide range of problems of a quantum mechanics and an optics. So far SO-FFT approach has been applied for the aim of the solution of one- and two-dimensional TDSE in the Cartesian coordinates \cite{FeitFleck1982}, three-dimensional TDSE in the spherical coordinates \cite{Fleck1988,Sorevik2009}, the time-dependent Kohn--Sham equations \cite{Castro2004}, the Gross--Pitaevskii equation \cite{Antoine2013}, and molecular dynamics \cite{Kosloff1988} as well. The method is non-effective in use for singular potentials \cite{Sorevik2009} because of both the inaccuracy of the Fourier transform approximation for the wavefunction with discontinuous derivative and the split operator method conditional stability. But the kind of problems characterized both by a smooth potential and a spectrum strictly bounded from above imply the Fourier transform to converge more rapidly \cite{Kosloff1988} by contrast to the finite difference schemes, the expansions over B-splines and finite elements, and any other polynomial approximation based approaches.
As compared to other time propagation techniques commonly used along with FFT, namely the leap-frog method and short iterative Lanczos method, the split-operator method appears to demand the less amount of memory for the service arrays storage \cite{Leforestier1991}. The latter is particularly important for the many-dimensional problems.

As a grid based approach SO-FFT may provide the solution only on a finite space region. If the wavefunction leaking beyond its boundaries becomes apparent during the evolution, then one encounters the unphysical reflection from the grid boundaries. Upon using SO-FFT the unphysical reflection is commonly suppressed \cite{Fleck1988} with the help of the adding of the complex absorbing potential (CAP) \cite{Muga2004}. However the CAP method exhibits a number of shortages. That are the complex potential absorbs best the small energy wavefunction components while the grid boundary is most fast reached by the high energy components, and also it distorts the wavefunction behavior in the CAP absent region \cite{Muga2004,Kosloff1986,Neuhauser1988,Manolopoulos2002}.

The CAP method disadvantages are absent in the exterior complex scaling (ECS) approach \cite{ReviewMcCurdy2004}. For the short iterative Arnoldi--Lanczos method the way for combining of the FFT and ECS has been demonstrated in \cite{Riss1998}. However the application of SO-FFT along with ECS is impeded by the explicit coordinate dependence of the scaled kinetic energy operator \cite{Leforestier1991}. In the present work a technique for the superposing SO-FFT and ECS is supposed. The technique is based upon the splitting of the scaled kinetic energy operator to the spatially homogeneous part (for which a step may be performed by means of FFT) and the absorbing term (which can be handled with the help of the finite difference). 
The approach under proposal has been tested through a calculation of continuum spectrum amplitudes for the model system by t\&E-SURFF method \cite{Serov2013}. For the purpose of the iterative solution of the stationary Schrodinger equation (SSE) (which is necessary for the accomplishing of the E-SURFF part of the t\&E-SURFF scheme) we suggest the FFT-based preconditioner.

The paper is organized as follows. In the Section \ref{sec:SOFT} the proposed method for the SO-FFT and ECS combining is exposed. The Section \ref{sec:Precond} describes the using of the preconditioned conjugated gradient method (PCG) with the aid of performing the SSE iterative solution, and also develops the advanced FFT-based preconditioner designed for the using with ECS. The section \ref{sec:Results} presents the SO-FFT testing, including the proposition of the advanced ECS contour.
 
Below we use the atomic units of measurement.

\section{Solution of the complex scaled TDSE by means of SO-FFT}\label{sec:SOFT}

Let us consider TDSE
\begin{eqnarray}
\frac{\partial\psi(x,t)}{\partial t}=\left[\hat{K}+U(x,t)\right]\psi(x,t),
\end{eqnarray}
where the kinetic energy operator
\begin{eqnarray}
\hat{K}&=&-\frac{1}{2}\frac{\partial^2}{\partial x^2}.
\end{eqnarray}
The formal solution of the TDSE is
\begin{eqnarray}
\psi(x,t+\tau)=e^{-i\left[\hat{K}+U\right]\tau}\psi(x,t).
\end{eqnarray}

The solution can be implemented by using constant-step discrete variable representation (DVR) \cite{Baye1995} over a space variable. Let us set up in the region $x\in [a,b]$ the uniform grid
\begin{eqnarray}
x_i=a+h(i-1/2),\quad i=1,\ldots,N,
\label{DVRgrid}
\end{eqnarray}
where $N=2^\nu$ is the nodes number ($\nu$ being an positive integer) and $h=(b-a)/N$ is the grid step.
We can expand $\psi(x,t)$ over the basis of Lagrange functions
\begin{eqnarray}
\xi_i(x) = \sqrt{h}\sum_{n=1}^N\varphi_n^*(x_i)\varphi_n(x).
\label{DVRbasis}
\end{eqnarray}
Here
\begin{eqnarray}
\varphi_n(x) = \frac{\exp(ik_nx)}{\sqrt{b-a}} \label{FFTbasis}
\end{eqnarray}
are the Fourier basis functions,
where the momentum $k_n$ is
\[
k_n=\left\{
 \begin{array}{ll}
 \frac{2\pi}{b-a}(n-1), & n\in [1,N/2]; \\
 -\frac{2\pi}{b-a}(N+1-n), & n\in [N/2+1,N].
 \end{array} \right.
\]
It is easy to check that Lagrange functions have property $\xi_i(x_j)=\sqrt{h}\delta_{ij}$, and coefficients of $\psi(x,t)$ expansion over Lagrange functions are $\psi_i(t)=\sqrt{h}\psi(x_i,t)$.

The evolution of the function $\bpsi(t)$ values vector at the grid nodes may be evaluated with the help of the split operator method along with FFT as follows:
\begin{eqnarray}
\bpsi(t+\tau)=\mx{F}^{-1}e^{-i\mx{K}\tau/2}\mx{F}e^{-i\mx{U}\tau}\mx{F}^{-1}e^{-i\mx{K}\tau/2}\mx{F}\bpsi(t) \label{SSFFT}
\end{eqnarray}
Here $\mx{F}$ and $\mx{F}^{-1}=\mx{F}^{\dag}$ are the orthogonal matrixes of the direct and the inverse discrete Fourier transforms (FT) respectively.
Their elements are defined by the expression
\begin{eqnarray}
F_{ni}=\varphi_n(x_i)\sqrt{h} \label{Fmatrix}
\end{eqnarray}
Due to the transition to the momentum representation through the direct FT and to the DVR through the inverse FT the matrixes of the potential $\mx{U}$ and the kinetic $\mx{K}$ energy are diagonal, that is $U_{ij}=U(x_i,t+\tau/2)\delta_{ij}$ and $K_{nm}=\frac{k_n^2}{2}\delta_{ij}$.
Thus the action of the propagators $e^{-i\mx{K}\tau}$ and $e^{-i\mx{U}\tau}$ reduces to the simple wavefunction multiplication by the phase factors. The multiplication by a fully populated matrix $\mx{F}$ generally requires $N^2$ computational operations. But it might be performed by using FFT that requires the number of operations of the order of $N\log_2N$.

In order to provide the outgoing wave boundary conditions one may use the ECS method consisting in the change of the real space variable to the complex one \cite{ReviewMcCurdy2004} in the following way:
\begin{eqnarray}
\frac{\partial\psi(z,t)}{\partial t}=\left[-\frac{1}{2}\frac{\partial^2}{\partial z^2}+U(z,t)\right]\psi(z,t)
\end{eqnarray}
Here the integration complex contour is conveniently described \cite{ReviewMcCurdy2004,Kurasov1994} by expression
\begin{eqnarray}
z(x)=\int_0^x q(x) dx. \label{ECSintdef}
\end{eqnarray}

Let us introduce the new function
\begin{eqnarray}
\tilde{\psi}(x)=\sqrt{q(x)}\psi(z(x)) \label{psi_zx}
\end{eqnarray}
and express the differentiation over the complex $z$ through the differentiation over the real $x$ as
\begin{eqnarray}
\frac{\partial\tilde{\psi}(x,t)}{\partial t}=\left\{-\frac{1}{2\sqrt{q(x)}}\frac{\partial}{\partial x}\frac{1}{q(x)}\frac{\partial}{\partial x}\frac{1}{\sqrt{q(x)}}+U(z(x),t)\right\}\tilde{\psi}(x,t).
\end{eqnarray}
It could be seen from here that upon the complex scaling performing the direct applying of FFT becomes impossible because the appearing of the explicit $x$ dependence in the kinetic energy operator
\begin{eqnarray}
\hat{K}_s=-\frac{1}{2\sqrt{q(x)}}\frac{\partial}{\partial x}\frac{1}{q(x)}\frac{\partial}{\partial x}\frac{1}{\sqrt{q(x)}}.
\end{eqnarray}
Hence the functions \eqref{FFTbasis} would not be its eigenfunctions anymore.

But we may divide the $\hat{K}_s$ operator into the two terms as $\hat{K}_s=\hat{K}+\hat{C}$, where
\begin{eqnarray}
\hat{C}&\equiv&\hat{K}_s-\hat{K}.
\end{eqnarray}
By construction $\hat{C} = 0$ in the domain where $q(x)=1$, i.e. outside the ECS region.
For the purpose of the operator $\hat{K}$ diagonalization one may use FFT, while the operator $\hat{C}$ (that actually provides the wavefunction absorbing in the scaling region) approximation can be performed through the finite-difference scheme since high accuracy calculation of the wavefunction is not necessarily needed in the ECS region.

We shall introduce the matrix $\mx{C}$ as the difference
\begin{eqnarray}
\mx{C}=\mx{K}_s-\mx{K}_\text{FD}, \label{C_FD}
\end{eqnarray}
where $\mx{K}_s$ and $\mx{K}_\text{FD}$ are respectively the operators $\hat{K}_s$ and $\hat{K}$ approximations implemented by the symmetric three-point difference scheme as
\begin{eqnarray}
[\mx{K}_s\bpsi]_{i}&=&-\frac{1}{2}\frac{q_i^{-1/2}}{h}\left[\frac{q_{i+1}^{-1/2}\psi_{i+1}-q_{i}^{-1/2}\psi_{i}}{q_{i+1/2}h}-\frac{q_{i}^{-1/2}\psi_{i}-q_{i-1}^{-1/2}\psi_{i-1}}{q_{i-1/2}h}\right]; \label{Ks_FD}
\\ {}
[\mx{K}_\text{FD}\bpsi]_{i}&=&-\frac{1}{2}\frac{\psi_{i+1}-2\psi_{i}+\psi_{i-1}}{h^2}. \label{K_FD}
\end{eqnarray}
This leads to a complex symmetric tridiagonal matrix $\mx{C}$. The matrix $\mx{C}$ symmetry is generally not crucial in a TDSE solution, but this property is needed for the providing of PCG convergence in a SSE solution process (that is considered in the next Section).

Thus the numerical scheme takes the following form:
\begin{eqnarray}
\bpsi(t+\tau)=\mx{F}^{-1}e^{-i\mx{K}\tau/2}\mx{F}e^{-i\mx{U}\tau/2}e^{-i\mx{C}\tau}e^{-i\mx{U}\tau/2}\mx{F}^{-1}e^{-i\mx{K}\tau/2}\mx{F}\bpsi(t),
\end{eqnarray}
where the multiplication by the exponential function $e^{-i\mx{C}\tau}$ of the tridiagonal matrix $\mx{C}$ may be performed with the help of the Cranck-Nicholson (CN) scheme as
\begin{eqnarray}
e^{-i\mx{C}\tau}\simeq \left(1+\frac{i\tau}{2}\mx{C}\right)^{-1}\left(1-\frac{i\tau}{2}\mx{C}\right).
\end{eqnarray}
CN requires the number of operations of the order of $N$.

\section{Solution of the stationary Schrodinger equation using FFT preconditioner}\label{sec:Precond}

The implementation of the E-SURFF approach \cite{McCurdy2007} and its generalization known as t\&E-SURFF \cite{Serov2013} both require the solution of the SSE with the right-hand side of the form
\begin{eqnarray}
\left[\hat{K}_s+U(z(x),t_{max})-E\right]\tilde\psi_E(x)=\tilde{f}(x)
\end{eqnarray}
Upon using the approximations described in the previous Section the operator in the left-hand side transforms into the matrix
\begin{eqnarray}
\mx{A}=\mx{F}^{-1}\mx{K}\mx{F}+\mx{C}+\mx{U}-E\mx{I}.
\end{eqnarray}
Here $\mx{I}$ is the identity matrix, $I_{ij}=\delta_{ij}$. The matrix $\mx{A}$ is a fully populated, so the direct solution of the equation
\begin{eqnarray}
\mx{A}\bpsi=\mx{f} \label{linsys}
\end{eqnarray}
by the Gaussian elimination method requires a number of operations of the order of $N^3$. However multiplication of a vector by the matrix $\mx{A}$ may be performed through $N\log_2N$ operations since the main overhead is caused by the multiplication by $\mx{F}^{-1}\mx{K}\mx{F}$. The multiplication by the tridiagonal matrix $\mx{C}$ as well as by the diagonal matrixes requires a number of operations of the order of $N$.

Since the matrix $\mx{A}$ is complex symmetric the system of equations \eqref{linsys} might be solved by means of an iteration method namely the preconditioned conjugated gradients (PCG) method \cite{GolubVanLoan,NumericalRecipes}. On every PCG iteration one performs the multiplication of a vector by the matrix $\mx{P}^{-1}\mx{A}$, where $\mx{P}$ is a preconditioner matrix that satisfies the equation $\mx{P}^{-1}\mx{A}\approx \mx{I}$ yet at the same time leads to a small operations number required for the equation $\mx{P}\mx{y}=\mx{r}$ solving.

In our case the natural choice is a preconditioner based on FFT in the following way
\begin{eqnarray}
\mx{P}^{-1}=\mx{F}^{-1}(\mx{K}-E\mx{I})^{-1}\mx{F} \label{PreconFFT}
\end{eqnarray}
where
\begin{eqnarray}
[(\mx{K}-E\mx{I})^{-1}]_{nm}=\left(\frac{k_n^2}{2}-E\right)^{-1}\delta_{nm}.
\end{eqnarray}
Upon utilizing the CAP approach (which is covered in details in the next Section) the simple preconditioner \eqref{PreconFFT} provides the fast PCG convergence. ECS test calculations showed that upon the condition $|q(x)|\sim 1$ PCG converges by rather small iterations number. But for the TFECS approach (see the next Section) $q(x)\to \infty$ near the grid boundaries that leads to the extremely slow convergence.

For this reason especially to deal with ECS we have developed the advanced FFT-based preconditioner as follows. Let us split the spectrum into the bands each containing the set of the momenta $k_n,\, n=n_l,\ldots,n_l+N_l-1$. Then let us write down an eigenvalue problem for the square matrix acting on the vector components in the one band only as follows
\begin{eqnarray}
\mx{K}_l\bs{\chi}_{\nu}=\lambda_{\nu}\bs{\chi}_{\nu}. \label{Prec_eigen_eq}
\end{eqnarray}
Here \( \chi_{\nu n}=0\) at \(n \notin [n_l,n_l+N_l-1] \), the solution number $\nu=n_l,\ldots,n_l+N_l-1$, and
\begin{eqnarray}
[\mx{K}_l]_{nm}=\left\{\begin{array}{ll}
 \frac{1}{2} k_nk_m Q_{n-n_l+1,m-n_l+1}, & n,m \in [n_l,n_l+N_l-1];\\
 0, & n,m \notin [n_l,n_l+N_l-1].\\
 \end{array}\right. \label{K_l}
\end{eqnarray}
In turn, here
\begin{eqnarray}
Q_{nm} = \int_{a}^{b} \varphi_n^*(x) [q(x)]^{-2} \varphi_m(x) dx.
\end{eqnarray}
Since $q(-x)=q(x)$, true is $Q_{mn}=Q_{nm}$, therefore the matrix $\mx{K}_l$ is complex symmetric.

The block diagonal matrix $\mx{\tilde{K}}=\sum_l\mx{K}_l$ may be considered as the matrix of an operator
\begin{eqnarray}
\hat{K}_a=-\frac{1}{2}\frac{\partial}{\partial x}\frac{1}{[q(x)]^2}\frac{\partial}{\partial x},
\end{eqnarray}
with artificially nullified off-blocks elements.
Using $\hat{K}_a$ instead of $\hat{K}_s$ here yields the following advantage. This choice allows to calculate $Q_{mn}$ elements only once for $1 \leq n,m \leq N_l$, then matrix $\mx{K}_l$ elements for any $n,m \in[1,N]$ are obtained from the $Q_{mn}$ through the simple multiplication by $k_nk_m/2$, according to \eqref{K_l}. The distortion of the eigenvalues and eigenvectors caused by the distinction of $\hat{K}_a$ from $\hat{K}_s$ are unessential by compare to the distortions because of the off-block elements nullifying.

Further let us construct the matrix $\mx{X}$ by using the equation \eqref{Prec_eigen_eq} solutions as the rows in the following way:
\begin{eqnarray}
X_{\nu m}=[\bs{\chi}_{\nu}]^T_m.
\end{eqnarray}
Since the matrix $\mx{\tilde{K}}$ is complex symmetric then the matrix $\mx{X}$ should be complex orthogonal that is $\mx{X}^{-1}=\mx{X}^T$.

The preconditioner we are proposing here is
\begin{eqnarray}
\mx{P}^{-1}=\mx{F}^{-1}\mx{X}^{-1}(\bs{\Lambda}-E\mx{I})^{-1}\mx{X}\mx{F}, \label{PreconFFT_ECS}
\end{eqnarray}
where $\bs{\Lambda}$ is a diagonal matrix with elements $\Lambda_{\nu\mu}=\lambda_{\nu}\delta_{\nu\mu}$, where $\lambda_{\nu}$ is an eigenvalue from \eqref{Prec_eigen_eq}. The matrix $\mx{X}$ is block-diagonal, therefore multiplication by it requires the operations number of the order of $N_lN$. Thus upon the condition $N_l\lesssim \log_2 N$ the solution of the preconditioner equation under proposal would not be essentially more time demanding than the one for the usual FFT preconditioner. At that the PCG convergence speeds up essentially due to the proximity of eigenvalues and eigenvectors of $\mx{P}$  for the ones of the matrix $\mx{A}$.

\begin{figure}[ht]
\includegraphics[angle=-90,width=0.45\columnwidth]{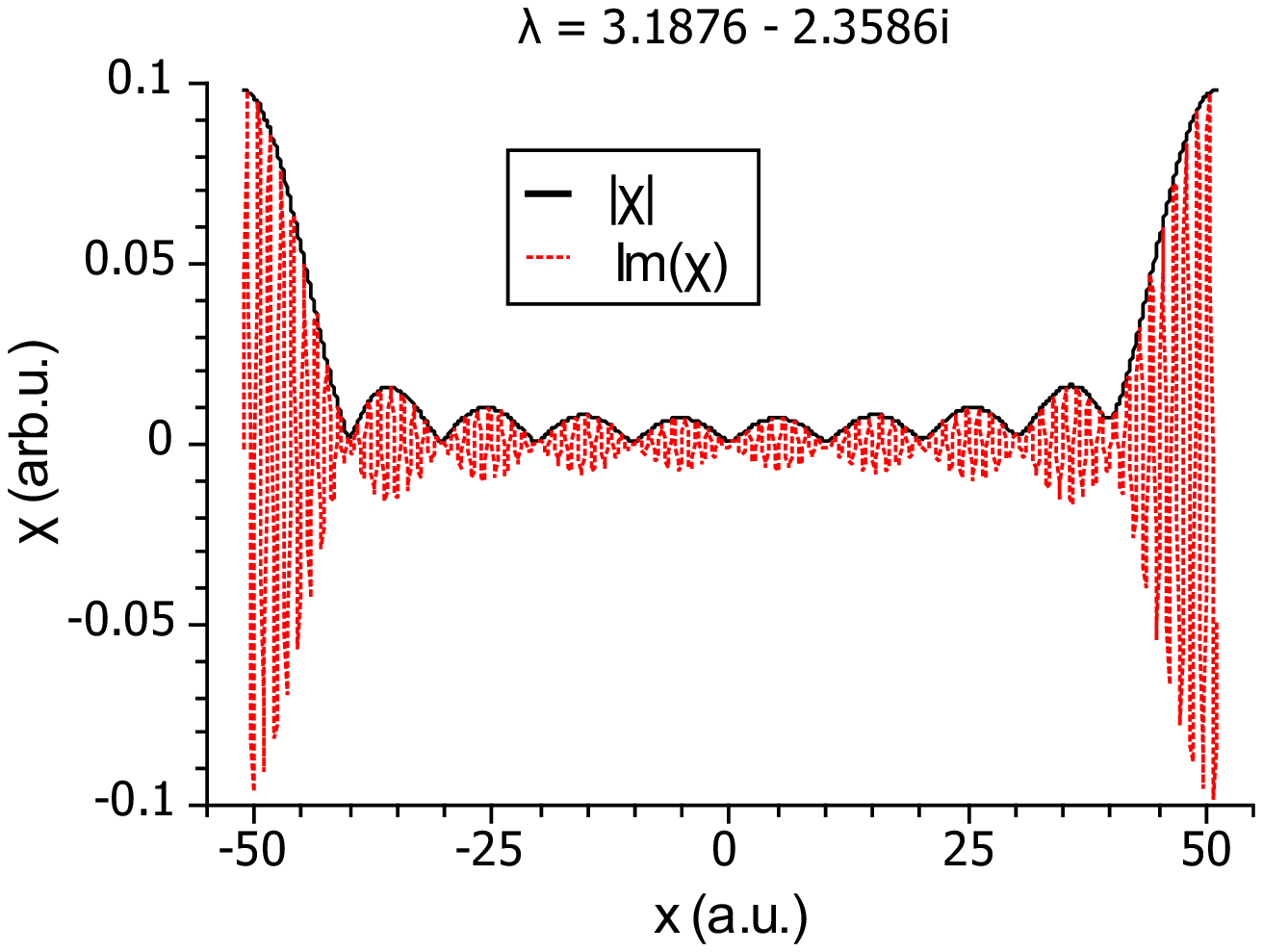}
\includegraphics[angle=-90,width=0.45\columnwidth]{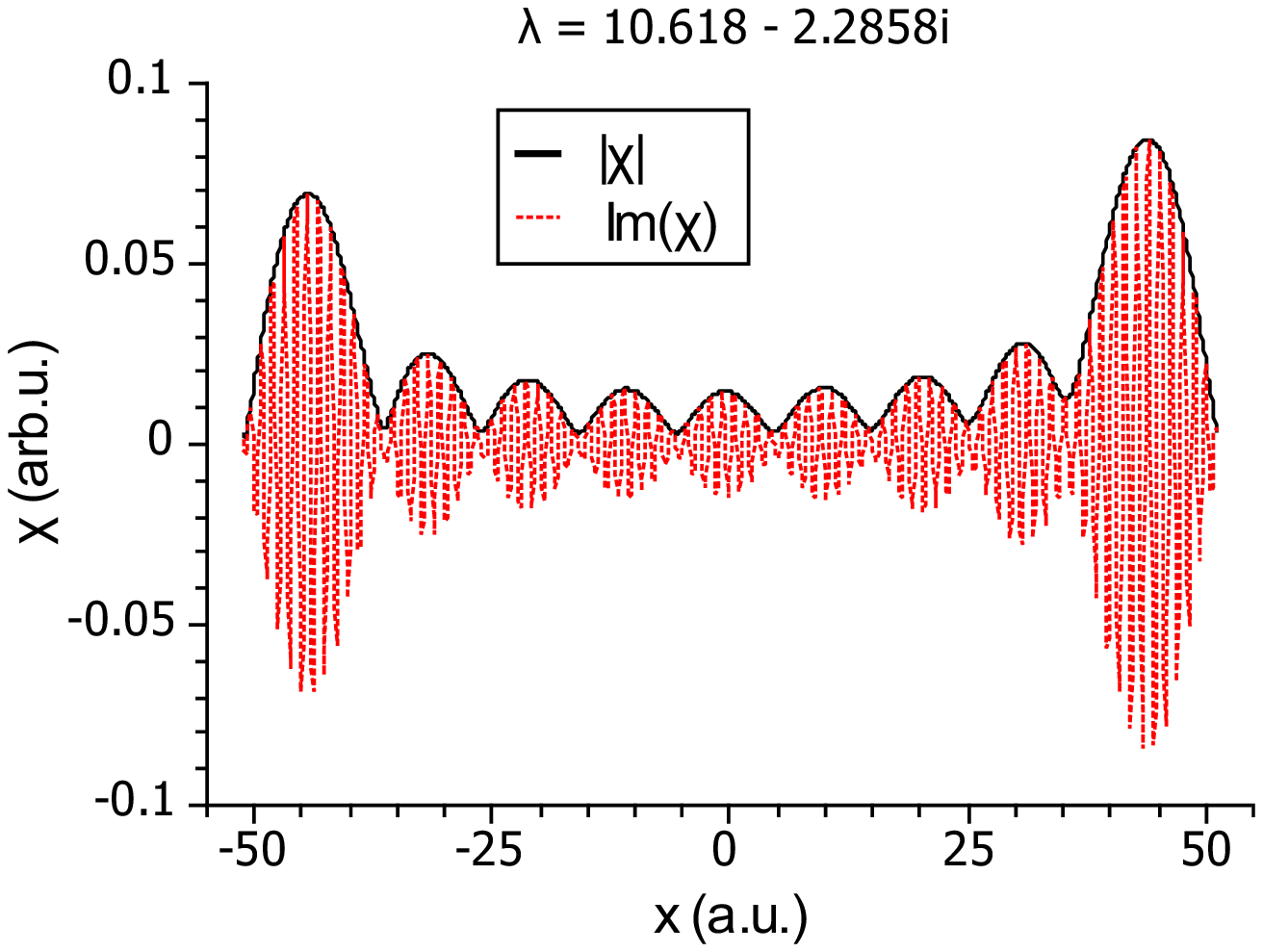}
\\
\includegraphics[angle=-90,width=0.45\columnwidth]{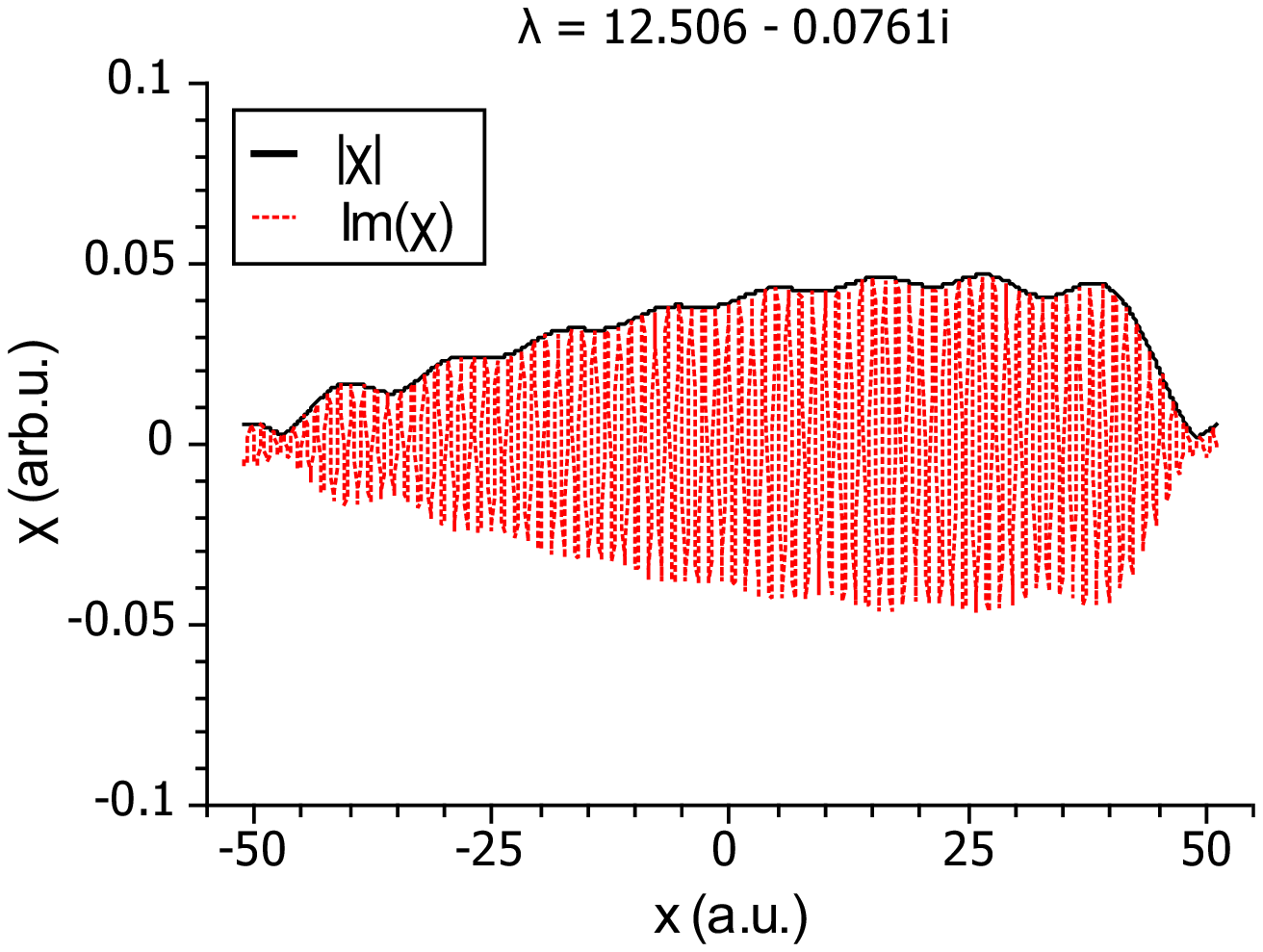}
\includegraphics[angle=-90,width=0.45\columnwidth]{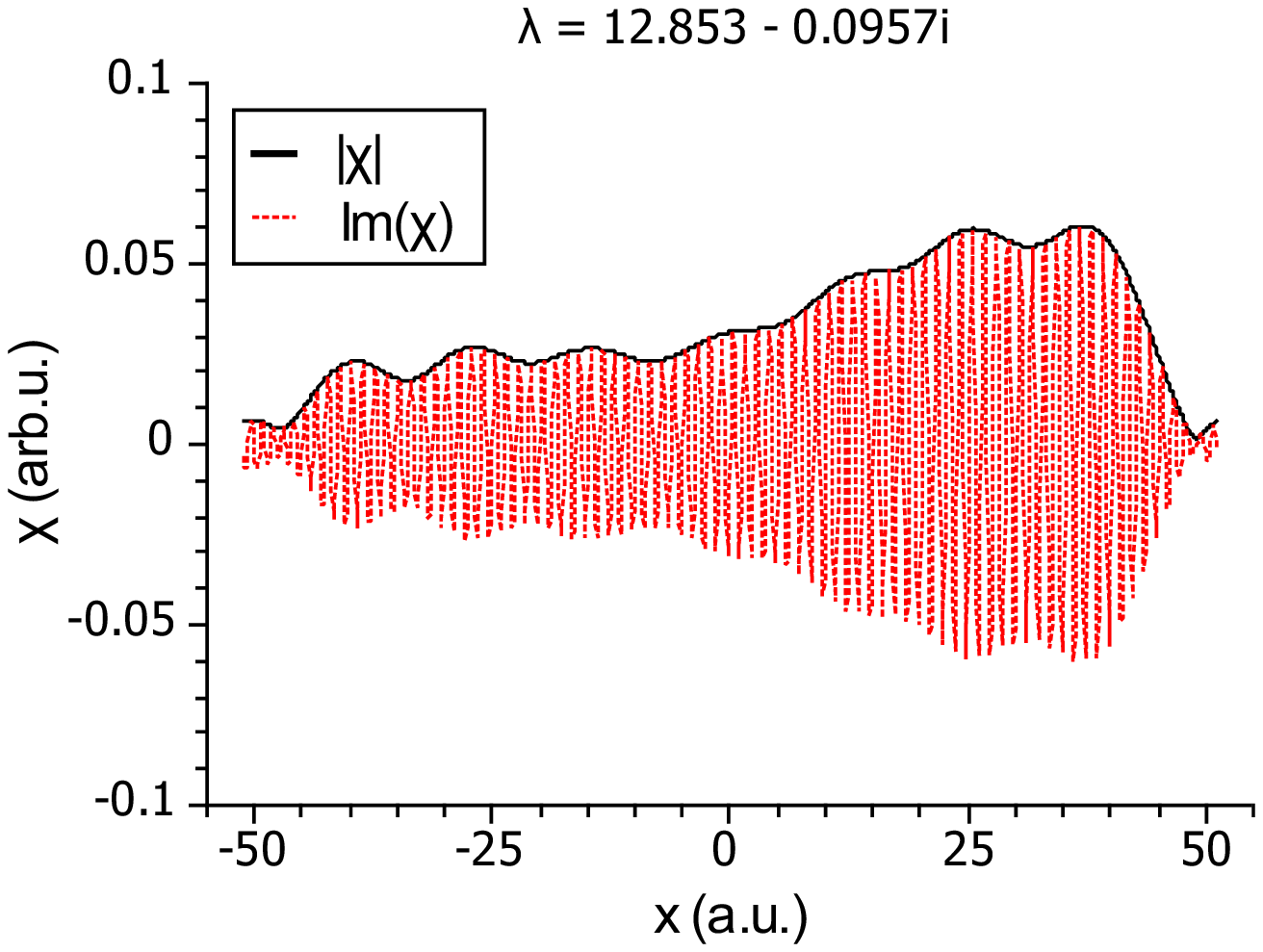}
\caption{Typical preconditioner basis functions with corresponding eigenvalues. Solid lines --- absolute value, dotted lines --- imaginary part.}
\label{fig:Precond}
\end{figure}
The Fig. \ref{fig:Precond} displays several typical preconditioner basis functions in the coordinate representation $\chi_{\nu}(x_i)=[\mx{F}^{-1}\bs{\chi}_{\nu}]_i$ for the case $N_l=10$.
These functions break down into the two kinds: the ones localized in the ECS range (upper row of Fig. \ref{fig:Precond}) and the ones diffused over the non-scaled region (lower row of Fig. \ref{fig:Precond}). The former typically have a large negative imaginary part of its eigenvalue. Since by construction the $\chi_{\nu}(x)$ have momentum uncertainty $\Delta k \leq \frac{2\pi}{b-a}N_l$ then the minimal region of its possible localization has the size $\Delta x \geq (b-a)/N_l$ due to the uncertainty principle. Thus one can draw a condition on the minimal $N_l$ needed for at least one function to be localized in the scaling region as follows: $N_l\geq (b-a)/2\Delta{x}_\text{CS}$, where $2\Delta{x}_\text{CS}$ is a joint width of the two scaling regions located at the grid edges (as we use the basis functions \eqref{FFTbasis} that are periodic, so these two regions may be considered as stitched together through the grid external boundaries).

It is not possible to split the spectrum into the bands of equal width $N_l=\text{const}$ in a $k_n$ symmetric way, while $k_1=0$ and $k_{N/2+1}$ has undefined sign, so the including of the corresponding components in any band leads to the disparity of the bands associated to the positive and negative $k_n$. Therefore we used the following splitting: two bands of the width $N_l=1$ each containing only one component, namely $k_1$ and $k_{N/2+1}$, and the set of constant width $N_l=N_\text{BW}=10$ bands, and finally the two mutually symmetric bands of the width $N_\text{BW}\leq N_l \leq 2N_\text{BW}-1$ which include the rest $k_n$ values on either side from $k_{N/2+1}$.

It should be noted that the FFT preconditioner (in both proposed variations) is easily generalized to the many-dimensional case. In \cite{ReviewMcCurdy2004,McCurdy2001} the solution of the many-dimensional SSE has been performed by means of the preconditioner based on the direct solution of the sparse block-tridiagonal matrix system. Authors of \cite{Serov2009} have used the preconditioner based on the orthogonal transformation to the representation of the one-dimensional problem eigenvectors set. Either of these two preconditioners utilizing implies carrying the number of operations of the order of $N^{3/2}$ (here $N$ being the full number of the two-dimensional grid points). FFT-based preconditioner needs the number of operations of the order of $N\log_2 N$ only for same case. 
It is inferior to the multigrid preconditioner \cite{Vanroose2012,Vanroose2014}, requiring as few as $N$ operations. However the FFT approach over-performs the multigrid approach when used for the problems for which the FFT approach exceeds the finite-difference schemes in the space approximation convergence rate \cite{Gholami2015}.

\section{Testing of the proposed methods}\label{sec:Results}

As a model example we have chosen the Gaussian packet
\begin{eqnarray}
\psi(x,0)=\pi^{-1/4}\exp\left(-\frac{x^2}{2}+iv_{0}x\right), \label{psi0}
\end{eqnarray}
moving with the speed  $v_{0}=2$ in the free space.
The Gaussian packet spectrum is
\begin{eqnarray}
A_{teor}(k)=\pi^{-1/4}\exp\left[-\frac{(k-v_{0})^2}{2}\right]. \label{Ateor}
\end{eqnarray}
So we can test the methods for the unphysical reflection suppression by comparing the theoretical spectrum to the one obtained by means of t\&E-SURFF.

\begin{figure}[ht]
	\centering
	\includegraphics[angle=-90,width=0.55\columnwidth]{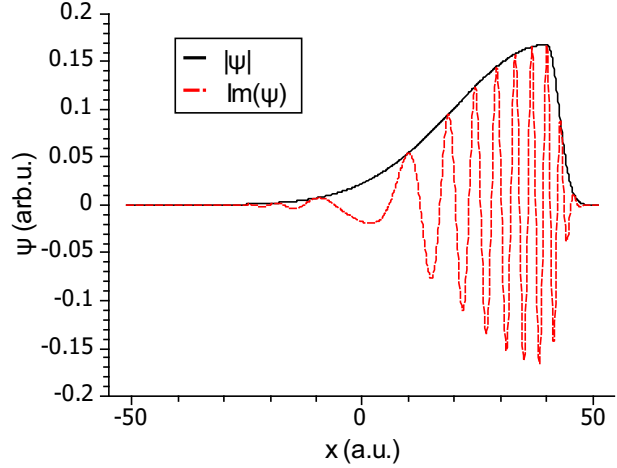}
	\caption{Wavefunction for $t=20$
	calculated by using SO-FFT with SECS. Solid line --- absolute value, dashed line --- imaginary part.}
	\label{fig:wf_ECS}
\end{figure}

For all the examples listed below we used the same grid parameters as follows. The grid boundaries were supposed to be $b=-a=r_{max}=51.2$, the ECS region was $|x|\geq r_{s} = 40$. The time step was taken to be equal to $\tau=h^2$, $h$ being the space grid step. The evolution has been performed up to the time $t_{max}=20$ (unless otherwised noted). At this $t_{max}$ the packet goes into the scaling region exactly by half (see the Fig.\ref{fig:wf_ECS}). Such a choice allows to estimate simultaneously the accuracy of the both t\&E-SURFF components, namely t-SURFF (which implies an amplitude extraction from the probability amplitude flux passed through the scaling region boundary) and E-SURFF (that extracts an amplitude from the rest function), since at such $t$ both components contribute equally to an amplitude under evaluation.

The ECS is usually introduced by the sharp contour rotation to the complex plane as follows:
\[
q(x)=\left\{
 \begin{array}{ll}
 1, & |x|<r_{s}; \\
 e^{i\theta_{s}}, & |x|>r_{s},
 \end{array} \right.
\]
where $\theta_{s}$ is the angle of the rotation to the complex plane, and $r_s$ is the distance from the origin to the rotation onset points.  However upon such a choice $\hat{C}$ has a discontinuity at the scaling region boundary that leads to the loss of the function $\tilde{\psi}(x)$ smoothness at the points $|x|=r_{s}$. For the non-smooth functions the Fourier expansion converges extremely slowly. It manifests itself in the strong wavefunction reflection from the scaling region.

\begin{figure}[ht]
\includegraphics[angle=-90,width=0.45\columnwidth]{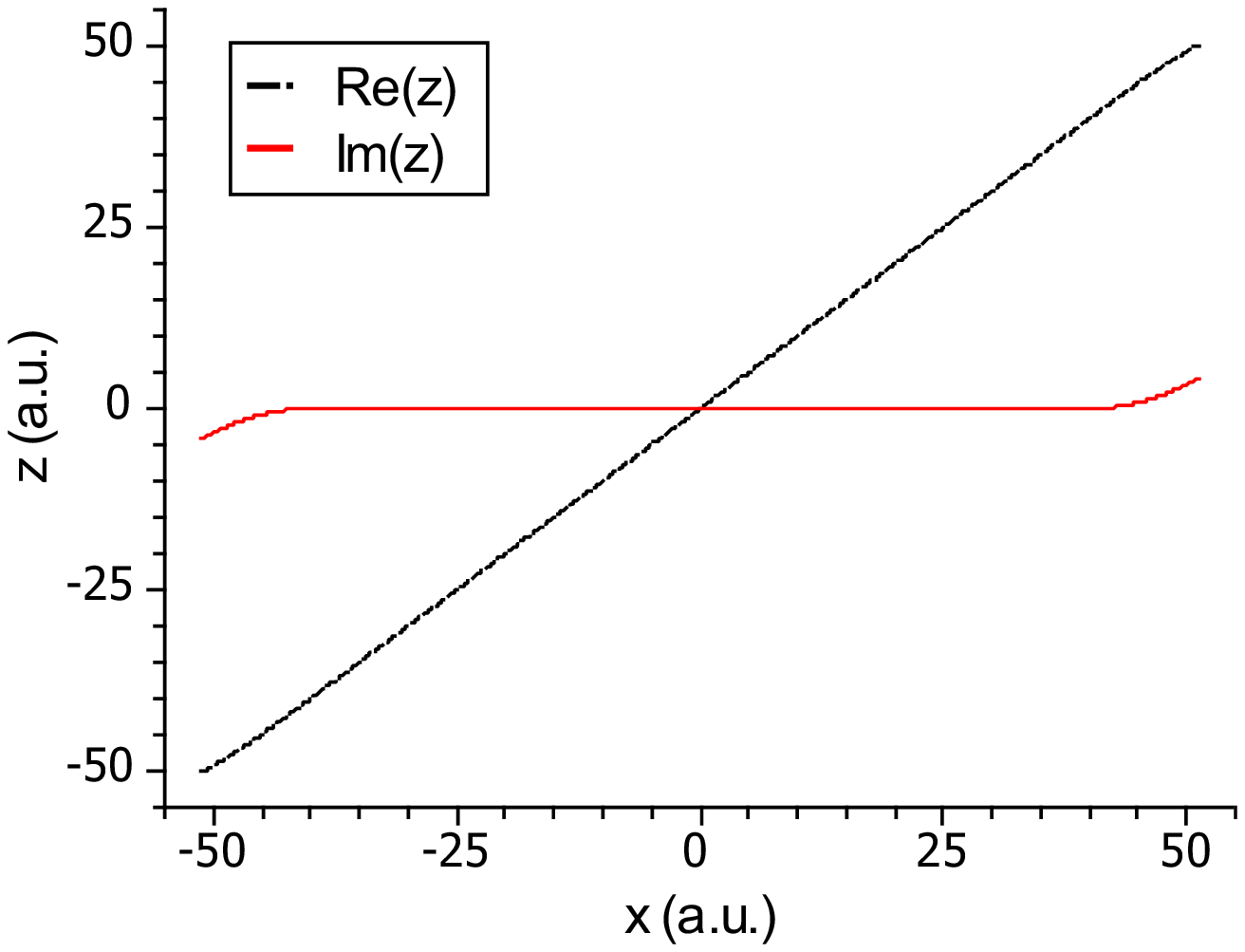}
\includegraphics[angle=-90,width=0.45\columnwidth]{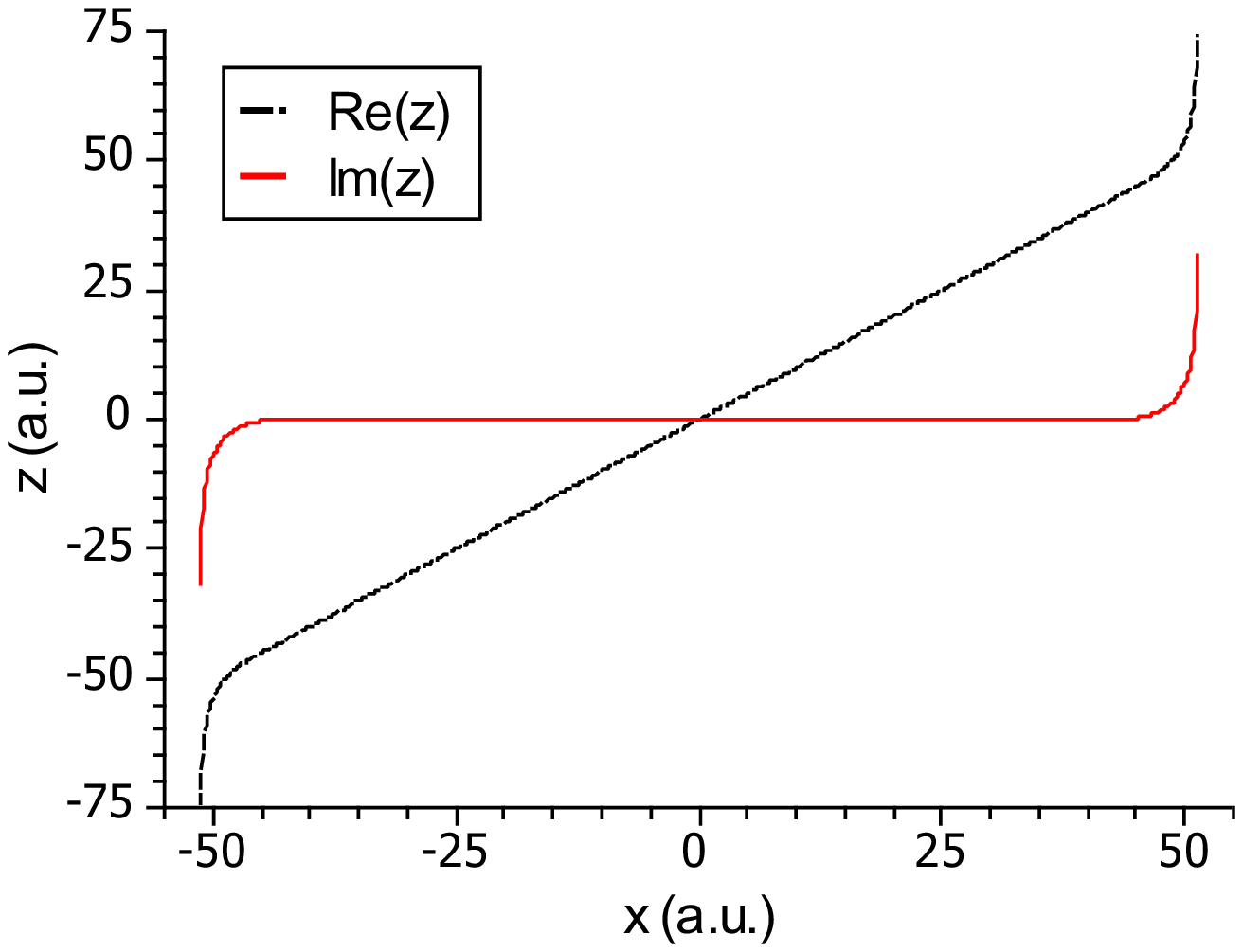}
\\
(a)\hspace{0.45\columnwidth}(b)
\caption{The ECS contours under consideration: left --- SECS, right --- TFECS. Solid lines --- imaginary part, dotted lines --- real part.}
\label{fig:ECS_contour}
\end{figure}

Therefore we have used the smooth ECS contour (smooth ECS, SECS) for which (see Fig. \ref{fig:ECS_contour}(a))
\begin{eqnarray}
q(|x|>r_{s})=\exp\left[\frac{i\pi}{4}\frac{|x|-r_{s}}{\Delta{x}_\text{CS}}\right], \label{qSECS}
\end{eqnarray}
where the scaling region width $\Delta{x}_\text{CS}=r_{max}-r_{s}$. For such contour, a maximal rotation angle is $\pi/4$ at $x=r_{max}$. 

\begin{figure}[ht]
\includegraphics[angle=-90,width=0.45\columnwidth]{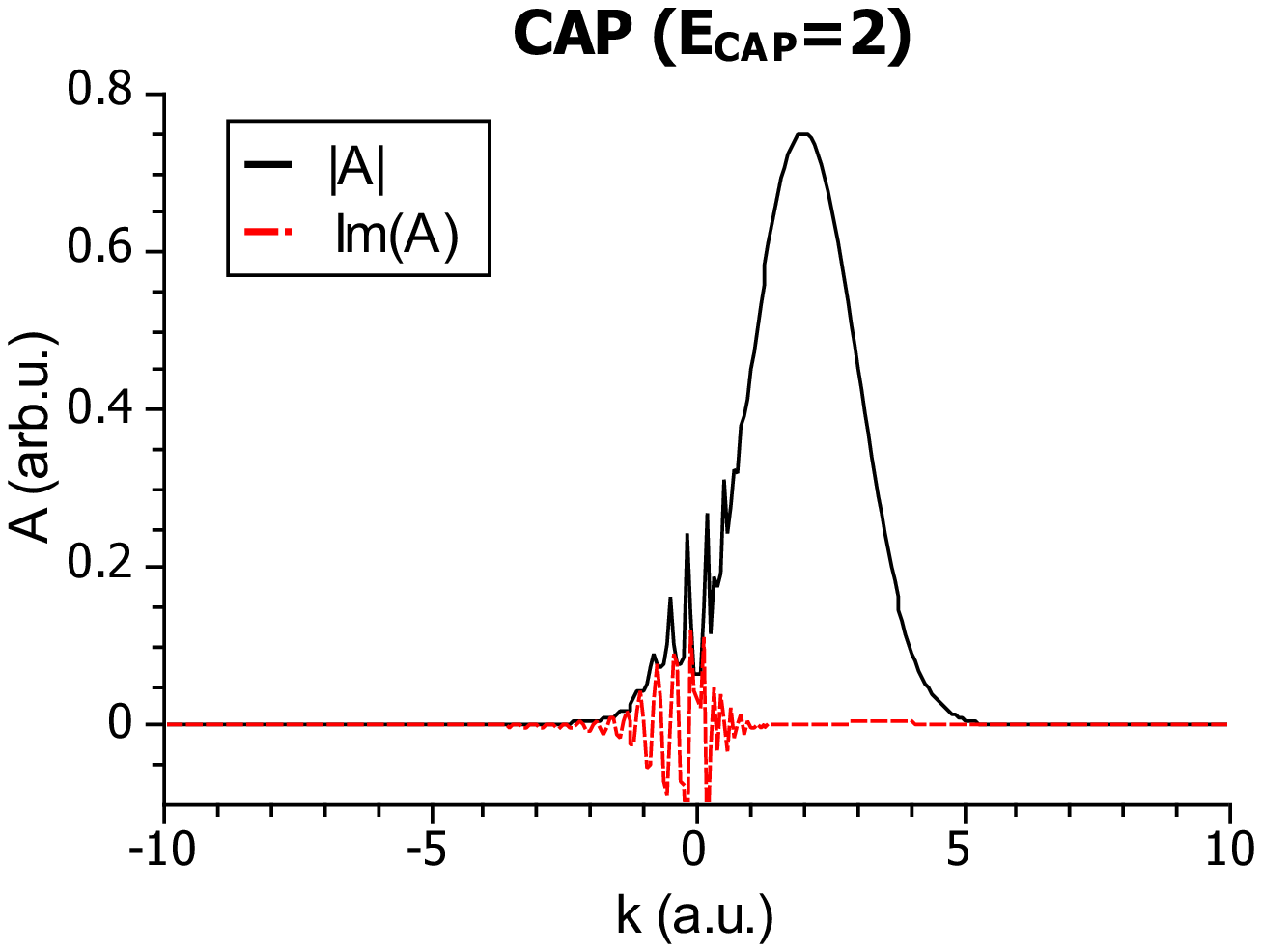}
\includegraphics[angle=-90,width=0.45\columnwidth]{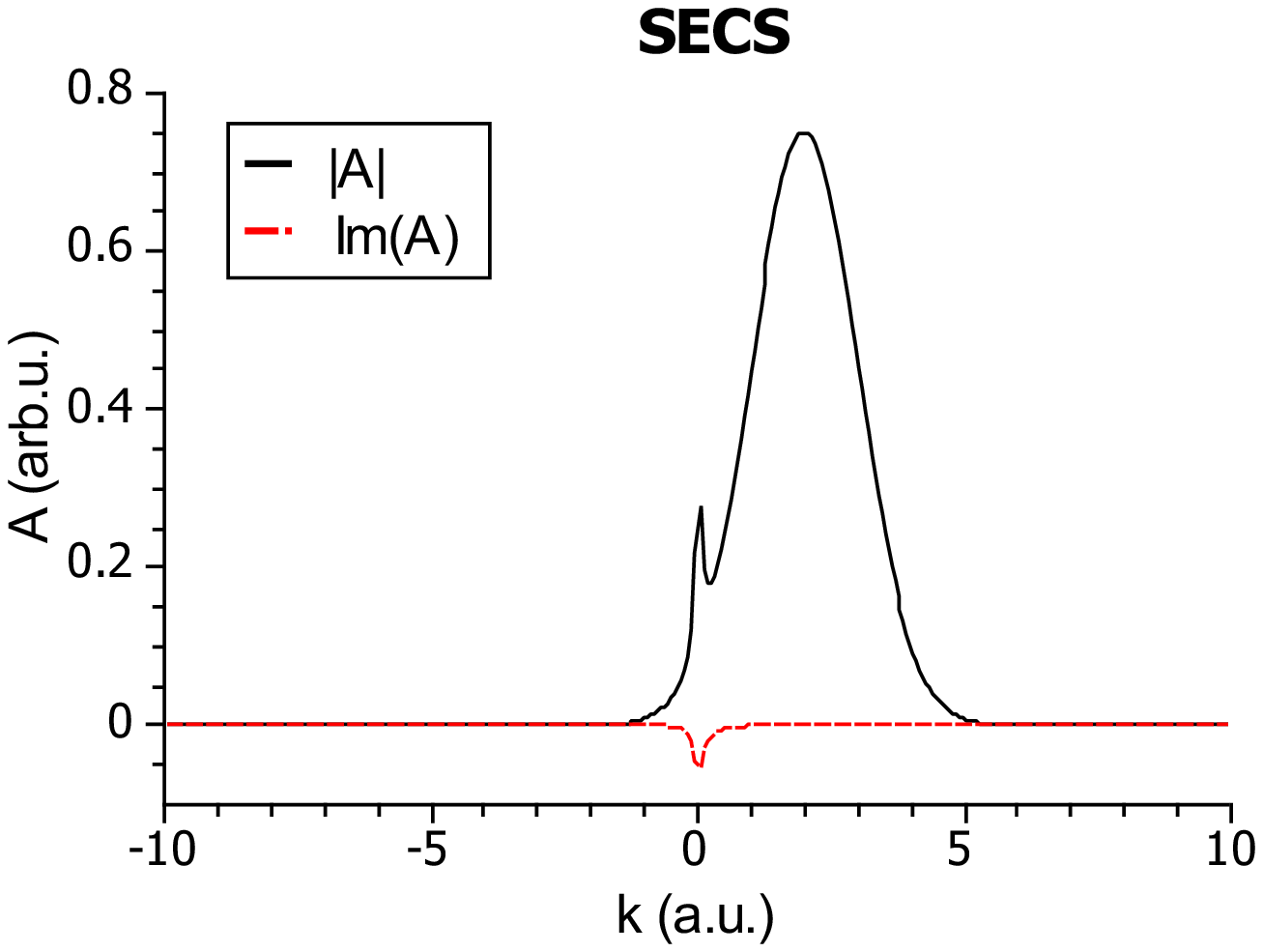}
\\
(a)\hspace{0.45\columnwidth}(b)
\\
\includegraphics[angle=-90,width=0.45\columnwidth]{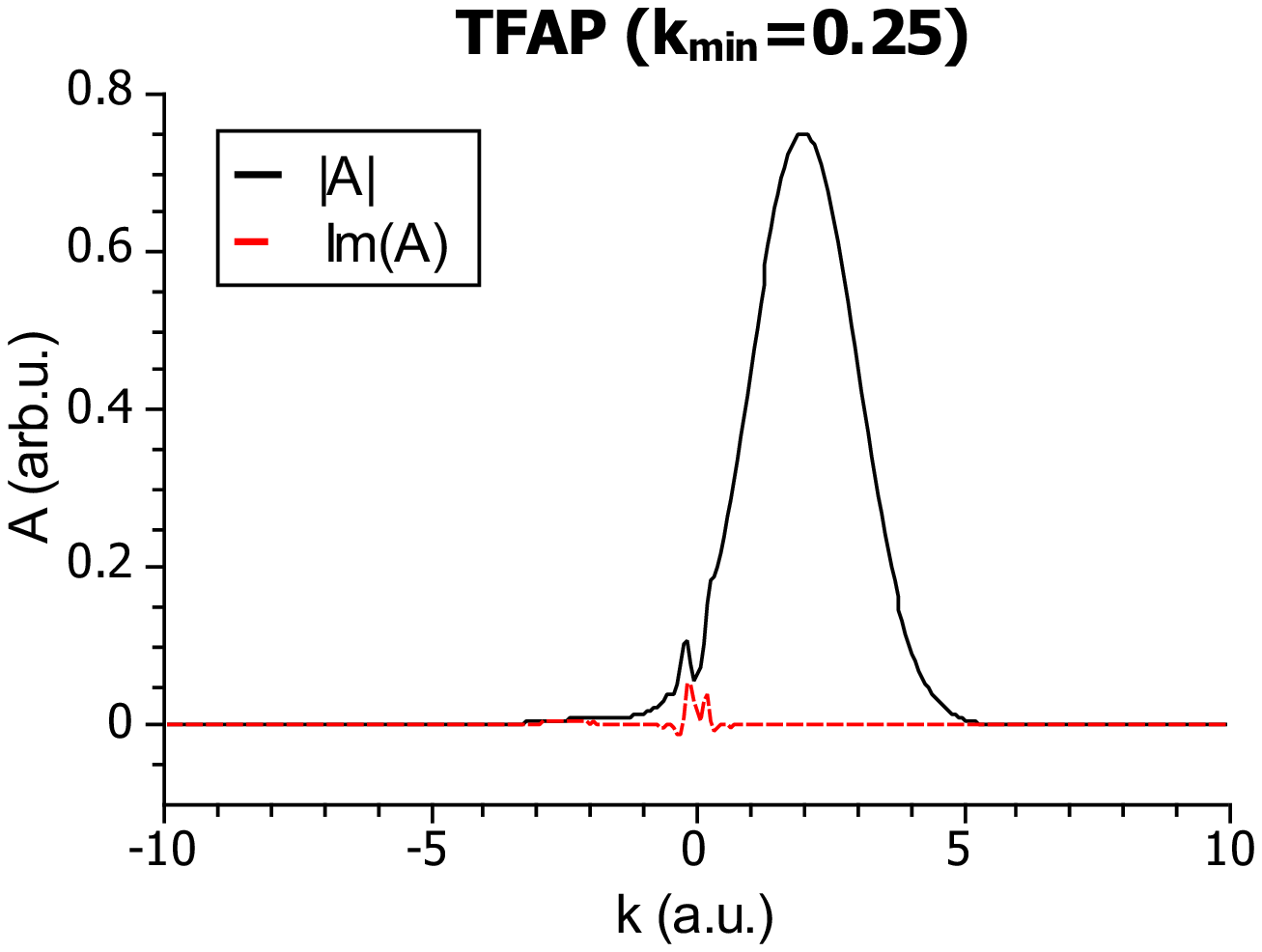}
\includegraphics[angle=-90,width=0.45\columnwidth]{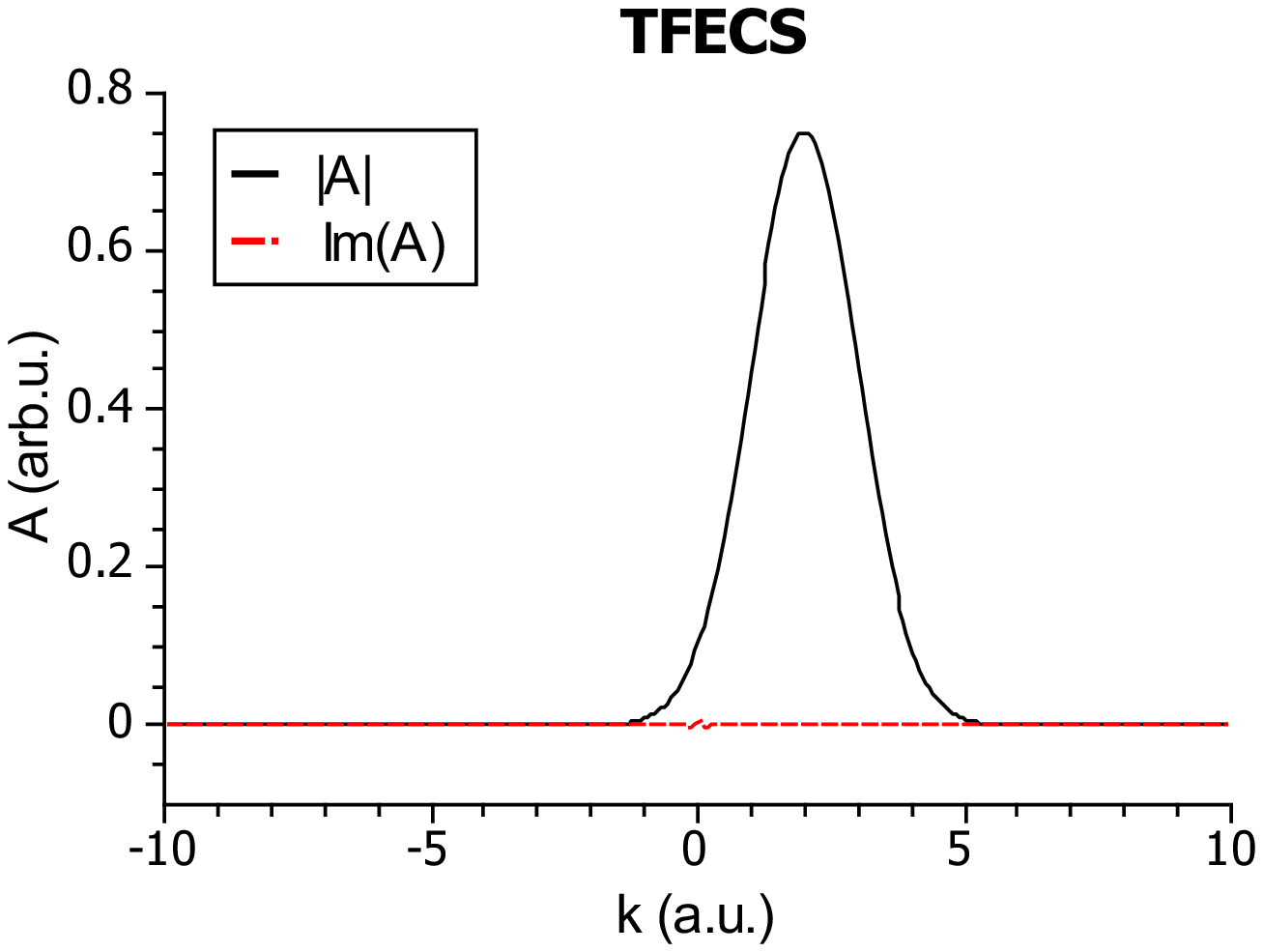}
\\
(c)\hspace{0.45\columnwidth}(d)
\caption{The amplitude $A(k)$ ($|A|$ --- solid lines, $\Im A$ --- dashed
 lines): a) CAP; b) SECS; c) TFAP; d) TFECS.}
\label{fig:CAP_vs_ECS}
\end{figure}

The result of the test amplitude evaluation by means of SECS for the points number $N=1024$ (corresponding to the grid step $h=0.1$) is shown in the Fig.\ref{fig:CAP_vs_ECS}(b). It is seen that the calculated curve differs the most from the Gaussian shape at the small momentum $k$ values. Under the ECS applying the boundary reflection coefficient has the form
\begin{align}
	R_\text{ECS}(E)=\exp\left[-2k\Im z(r_{max})\right]. \label{Rcoeff_ECS}
\end{align}
Its tending to 1 at $k\to 0$ results in the absence of the unphysical reflection suppression for the small $k$, therefore the situation observed was expected.

In order to examine the correctness of the phase evaluation we represent in the Fig.\ref{fig:CAP_vs_ECS} also the amplitude imaginary part that should be equal to zero due to \eqref{Ateor} for the model system under consideration.

The authors who utilize SO-FFT commonly use the complex absorbing potential (CAP) for suppression of the unphysical reflection. It makes sense to compare the CAP and ECS efficiencies. As it is shown in \cite{Riss1998}, for the fixed kinetic energy value $E=E_\text{CAP}$ the ECS solution is equivalent to the solution with the complex absorbing potential of the form
\begin{eqnarray}
U_a(x)=E_\text{CAP}[1-q(x)^2]. \label{CAP_via_q}
\end{eqnarray}

As a first CAP version we used the potential given by the formula \eqref{CAP_via_q} for $q(x)$ from \eqref{qSECS} at $E_\text{CAP}=2$ a.u., that is the energy corresponding to the amplitude maximum. Such CAP is close to the commonly used linear CAP. The result is presented in the Fig.\ref{fig:CAP_vs_ECS}(a). One can easily see the essential curve distortions exceeding those for SECS (Fig. \ref{fig:CAP_vs_ECS}(b)). By virtue of the quasiclassical approximation it is not difficult to derive the formula for the grid boundary reflection under the CAP action:
\begin{align}
	R_\text{CAP}(E) \simeq \exp\left[ \frac{2}{k} \int_{r_{s}}^{r_{max}} \Im U_{a}(x) dx\right] \label{Rcoeff_CAP}
\end{align}
That is the maximum reflection is to occur at the large $k$ values. As the greatest distortions in the Fig. \ref{fig:CAP_vs_ECS}(a) are observed at small $k$ values, they are expected to be caused not by the grid boundary reflection, but by the reflection from the CAP itself.

In the work \cite{Manolopoulos2002} an advanced CAP version is proposed, namely so-called transmission-free absorbing potential (TFAP) having the form
\begin{eqnarray}
U_a(|x|>r_{s})=-i\frac{k_\text{min}^2}{2}y\left(\frac{|x|-r_{s}}{\Delta{x}_\text{CS}}\right), \label{TFAP}
\end{eqnarray}
where
\begin{eqnarray}
y\left(s\right)=\frac{1}{c^2}\left[as-bs^3+\frac{4}{(1-s)^2}-\frac{4}{(1+s)^2}\right], \label{yTFAP}
\end{eqnarray}
and parameters are $a=c^3-16$, $b=c^3-17$, $c \approx 2.622057$. This potential was obtained by the authors of \cite{Manolopoulos2002} from the condition of minimization of the wavefunction deviation from the quasiclassical outgoing wave hence the minimization of the reflection from the potential itself. The parameter $k_\text{min}$ in \eqref{TFAP} has a meaning of the smallest momentum needed for the reflection from the potential to be small. Upon $k_{min}$ value increasing the reflection from the potential grows, and upon its decreasing one encounters the reflection from the grid boundary since the absorbing potential weakens at small $k_{min}$. The Fig. \ref{fig:CAP_vs_ECS}(c) displays the results for $k_{min}=0.25$ a.u. It is apparent that the results are much more accurate than those obtained by the linear CAP.

By analogy with TFAP we propose transmission-free ECS (TFECS) approach making use of the following contour:
\begin{eqnarray}
q(|x|>r_{s})=\sqrt{1+i\tilde{y}\left(\frac{|x|-r_{s}}{\Delta{x}_\text{CS}}\right)}, \label{qTFECS}
\end{eqnarray}
where
\begin{eqnarray}
\tilde{y}\left(s\right)=\alpha\left[\frac{1}{(1-s)^2}-\frac{1}{(1+s)^2}-4s\right].
\end{eqnarray}
That is we chose $q(x)$ meaning that is equivalent to the absorbing potential of the form $U_a(x)=-i\tilde{y}\left[|x-r_{s}|/\Delta{x}_\text{CS}\right]$ due to \eqref{CAP_via_q}. The function $\tilde{y}(s)$ asymptotic behavior at $s\to 1$ ($x\to r_{max}$) is analogous to that for $y(s)$ from \eqref{yTFAP}, however $\tilde{y}(s)\sim s^3$ at $s\to 0$ ($x\to r_{s}$) while $y(s)\sim s$.  The reasons for such a choice will be described below. At such $q(x)$ the complex coordinate $z(x)$ tends to the infinity logarithmically at $x\to r_{max}$. But by virtue of the grid $x_i$ shift by $h/2$ the value $z(x_N)=-z(x_1)\simeq\sqrt{\alpha i}\Delta{x}_\text{CS}\ln(2\Delta{x}_\text{CS}/h)$ becomes finite (see Fig. \ref{fig:ECS_contour}(b)). This is the reason for the choice of Eq.\eqref{DVRgrid} for grid definition. The Fig. \ref{fig:CAP_vs_ECS}(d) represents the TFECS results for the parameter $\alpha=0.5$. It is seen that TFECS yields much more accurate results than those obtained by the rest methods under consideration.

Upon the $\Im z(x_N)$ value substitution in the Eq.\eqref{Rcoeff_ECS}, one may evaluate the coefficient of the grid boundary reflection under TFECS using in the following way:
\begin{align}
	R_\text{TFECS}(E)\simeq \left(\frac{h}{2\Delta{x}_\text{CS}}\right)^{\sqrt{2\alpha}k\Delta{x}_\text{CS}}. \label{Rcoeff_TFECS}
\end{align}

\begin{figure}
	\centering
		\includegraphics[angle=-90,width=0.55\columnwidth]{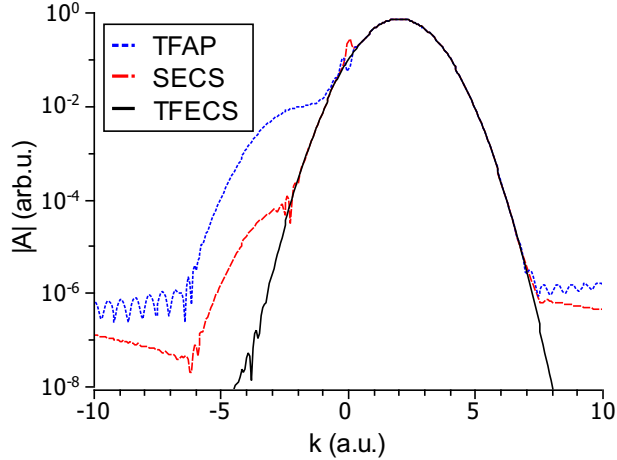}
	\caption{The absolute value of the amplitude $A(k)$: TFAP (dashed line), SECS (dotted line), and TFECS (solid line).}
	\label{fig:Alog}
\end{figure}

For some physical problems the calculation accuracy is crucial at large $k$ values. Therefore we shall consider this issue.
The Fig.\ref{fig:Alog} shows the results obtained by TFAP, SECS, and TFECS in the logarithmic scale. Since the wavepacket moves to the right, the reflection from the TFAP/ECS region produces a packet moving to the left that results in the amplitude distortion at $k<0$. It is apparent that even the simple SECS suppresses the unphysical reflection much better than TFAP does. Nevertheless at $k<-2.5$ one can easily discern the error of the order of 10$^{-4}$ entailed by the reflection. As the ``exact'' complex scaling should not cause any reflection at large $|k|$, so the main reason for the results to deviate from the theoretical form at large $|k|$ appears to be the inaccuracy of the finite-difference approximation of the absorbing operator $\hat{C}$ that leads to a small parasite potential from which the reflection does occur. The greatest error arises in the points of the function $q(x)$ higher derivatives discontinuity, namely the point of the rotation to the complex plane, $x=r_{s}$. 

As under the TFECS using the function $q(x)$ has the derivative discontinuity of the fourth order only near this point, so TFECS applying results in the reflection from the error-caused parasite potential yielding the amplitude error of the order of 10$^{-6}$ only (that corresponds to the cross section error $\sim 10^{-12}$). Thus upon the TFECS utilizing $\Im z(r_{max})$ provides the essential unphysical boundary reflection suppression, while the smoothness of the complex plane rotation ensures the smallness of the error caused by the absorbing operator $\hat{C}$ approximation by the finite difference scheme.

\begin{figure}[ht]
\includegraphics[angle=-90,width=0.45\columnwidth]{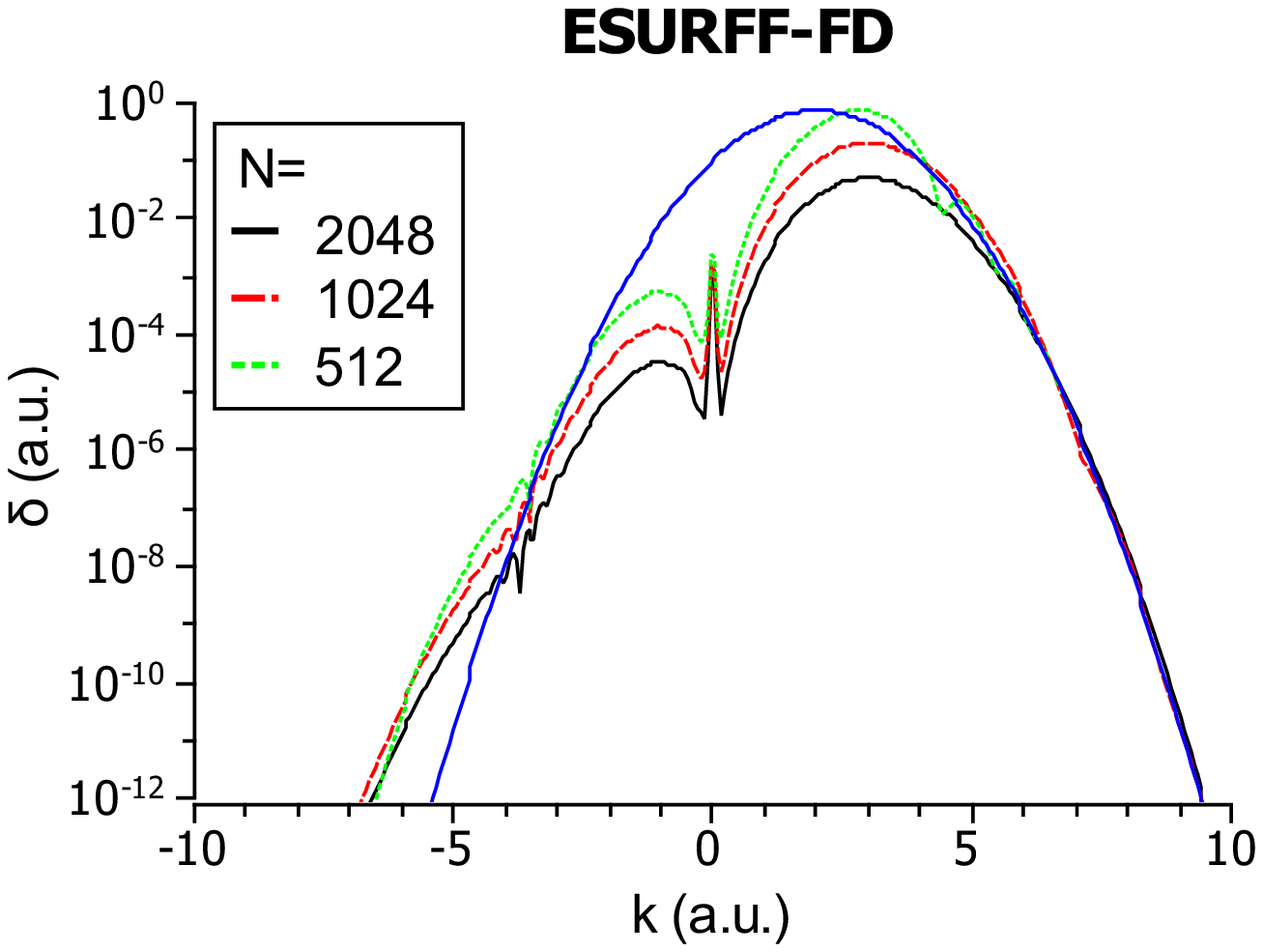}
\includegraphics[angle=-90,width=0.45\columnwidth]{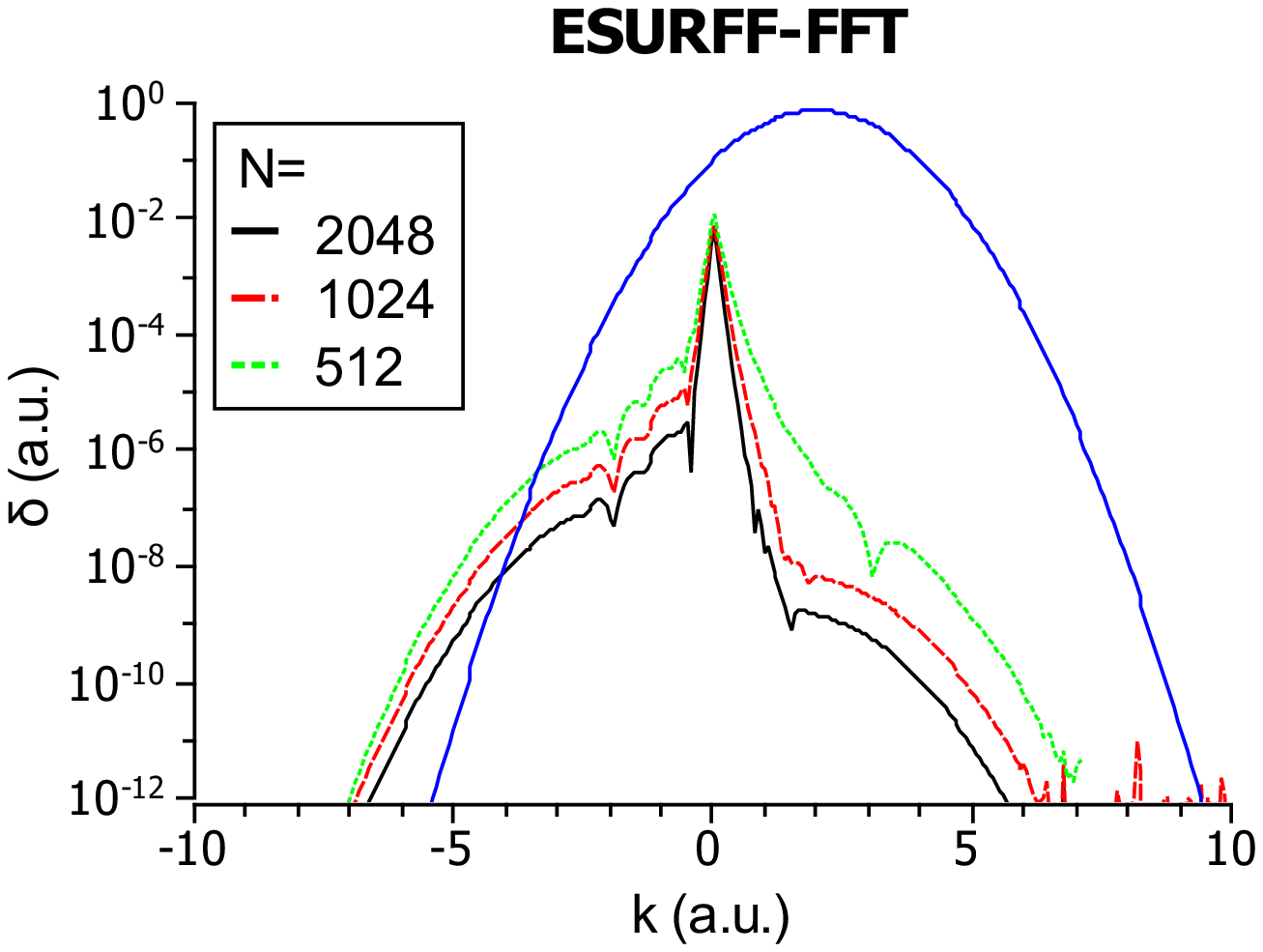}
\\
(a)\hspace{0.45\columnwidth}(b)
\caption{The error $\delta=|A(k)-A_\text{teor}(k)|$ calculated by using E-SURFF (from the wavefunction for $t=0$) for the different values of the grid points number: $N=512$ (dotted line), $N=1024$ (dashed line), $N=2048$ (solid line). The dependence $|A_\text{teor}(k)|$ is also shown (thin solid line). Left --- FD, right --- FFT.}
\label{fig:h_convergence}
\end{figure}

Further let us examine more thoroughly the solution convergence to the exact one with the grid step $h$ decreasing, hereinafter under the TFECS using. Since for the aim of ECS introduction to SO-FFT we use the three-point
finite-difference scheme, it makes sense to compare our approach accuracy to that obtained by the three-point finite-difference scheme \eqref{Ks_FD} for the full $\hat{K}_s$. We shall use ``pure'' E-SURFF (that is the amplitude extraction from the wavefunction at $t=0$) in order to separate the spatial approximation error effect. The Fig.\ref{fig:h_convergence}(a) presents the error $\delta(k)=|A(k)-A_\text{teor}(k)|$ of the amplitude obtained by the finite-difference scheme, for the spatial step values $h=$ 0.2, 0.1, и 0.05.
In order to demonstrate the relative value of the error,
we display on the same figure the theoretical curve $|A_\text{teor}(k)|$. One can easily see the error largeness.  As expected for the second-order scheme, the error decreases by the factor of 4 upon the step twofold decreasing.

The Fig.\ref{fig:h_convergence}(b) represents the result obtained with the help of FFT. It is apparent that the error is many orders smaller than that given by FD (except for the one for the smallest values $k$). The main error is entailed by the parasite reflection arising due to the inaccuracy of the finite-difference $\hat{C}$ approximation. Since we use the finite-difference scheme of the second order, the coefficient of the parasite reflection should depend on $h$ quadratically, exactly as it is observed in the Fig.\ref{fig:h_convergence}(b). Though formally the convergence rate is the same as upon the using of the finite-difference scheme for the full $\hat{K}_s$, the absolute value of the error is much smaller. 

The peak of the error at $k=0$ arises due to the reflection from the grid boundary. Therefore, this peak is hardly
reduced with decreasing of $h$. It may be suppressed significantly by the scaling region $\Delta{x}_\text{CS}$ increasing only, as it follows from Eq.\eqref{Rcoeff_TFECS}.

\begin{figure}[ht]
\includegraphics[angle=-90,width=0.45\columnwidth]{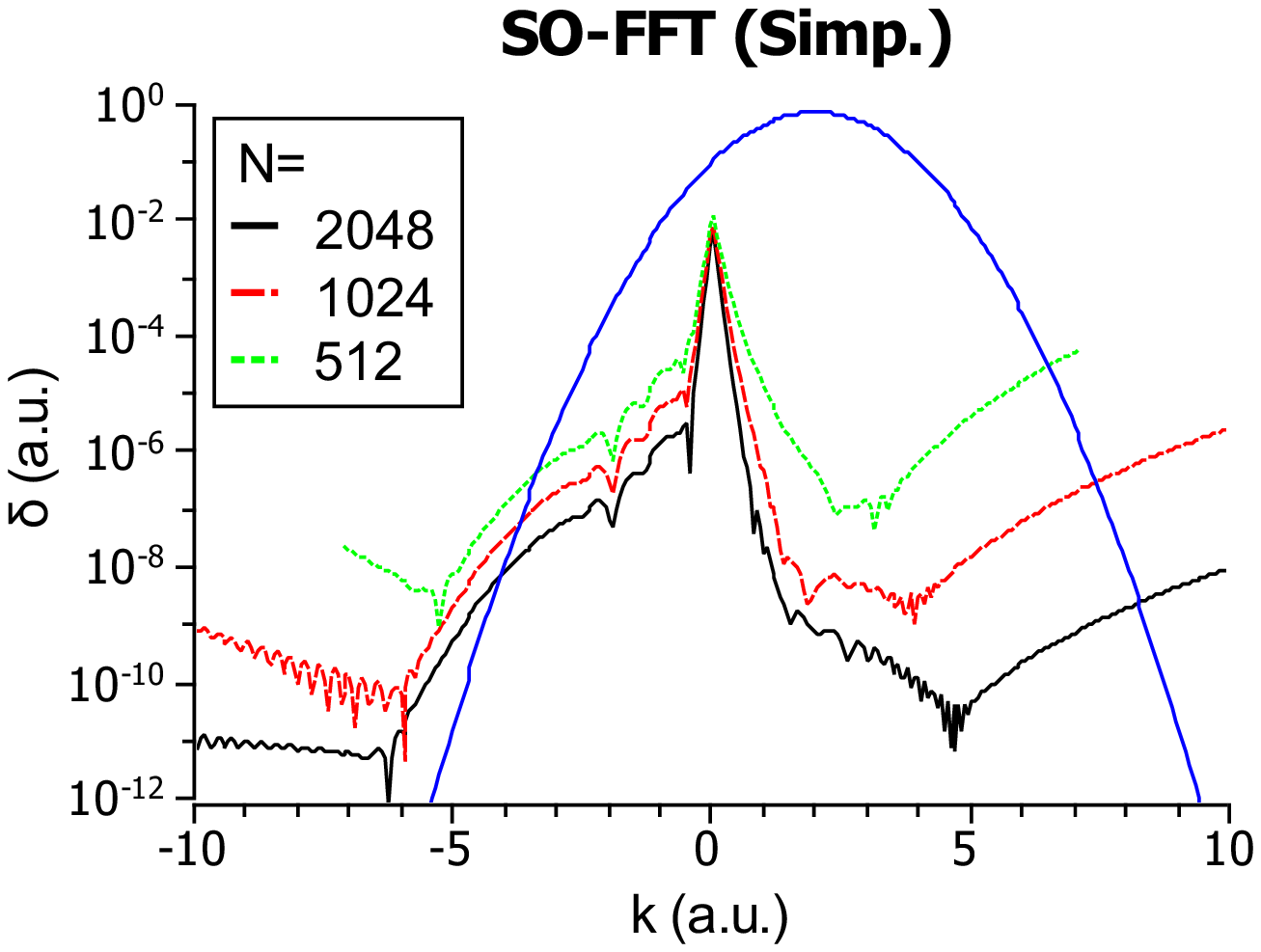}
\includegraphics[angle=-90,width=0.45\columnwidth]{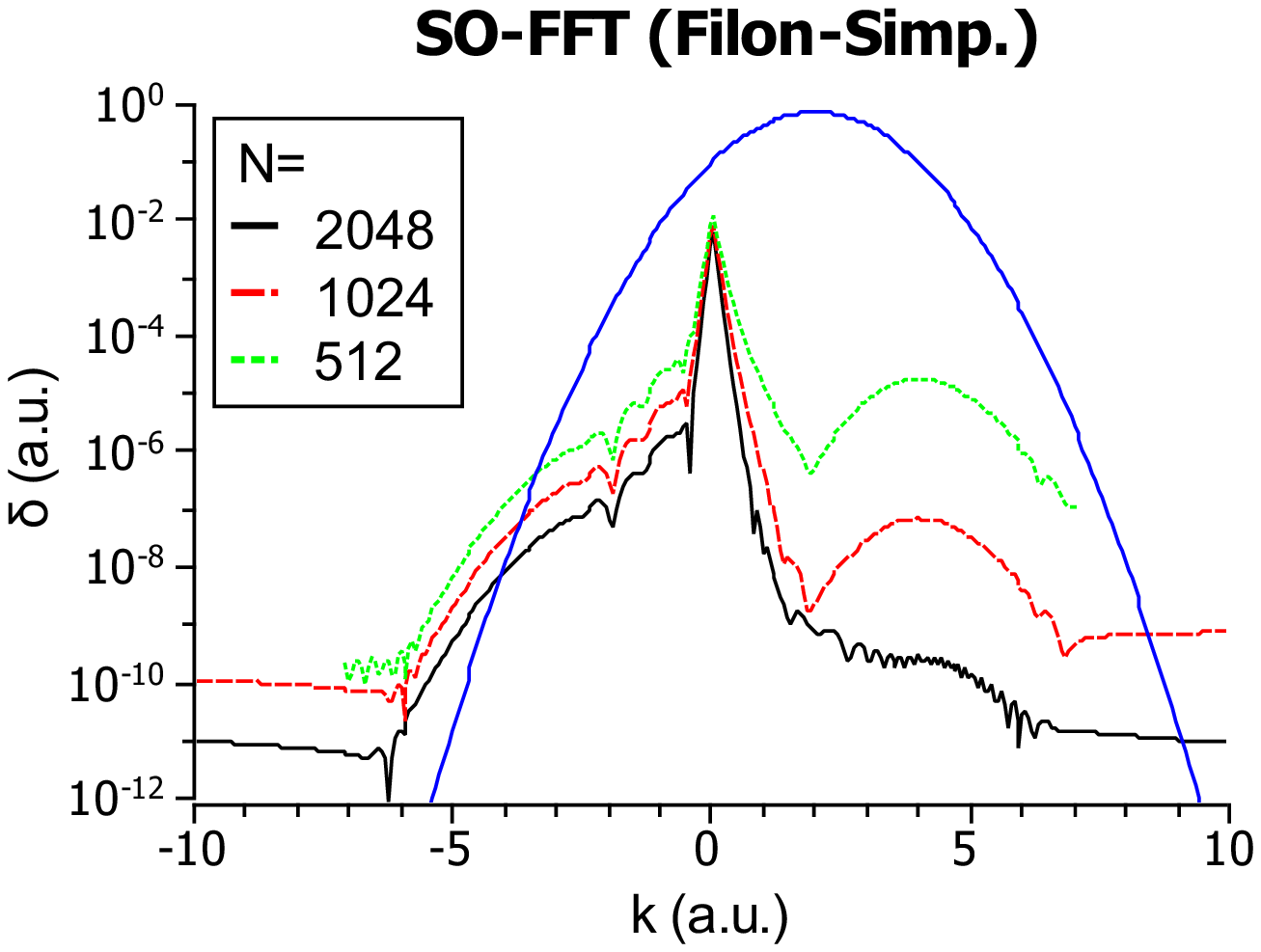}
\\
(a)\hspace{0.45\columnwidth}(b)
\caption{The error $\delta=|A(k)-A_\text{teor}(k)|$ calculated by using tE-SURFF (from wavefunction for $t=20$) for the different values of the grid points number: $N=512$ (dotted line), $N=1024$ (dashed line), $N=2048$ (solid line), with the time integration performed by Simpson's rule (left) and with Filon-Simpson's rule (right). The dependence $|A_\text{teor}(k)|$ is also shown (thin solid line).}
\label{fig:h_convergence_t20}
\end{figure}

The Fig. \ref{fig:h_convergence_t20} shows $\delta(k)$ for SO-FFT at $t_{max}=20$ a.u. The error increasing as compared to that of E-SURFF (Fig.\ref{fig:h_convergence}(b)) is mainly caused by the error of the approximate integration rule for the time integration of the probability amplitude flux \cite{Serov2013}. In the Fig. \ref{fig:h_convergence_t20}(a) one can see the error appearing upon the using of Simpson's rule for the time integration of the probability amplitude flux. At $k>5$ a.u. the integration error dominates the other error sources. Since we used the time step $\tau=h^2$ whereas the Simpson's rule error is $O(\tau^4)$, so the integration error depends on the grid step $h$ as $O(h^8)$, as is observed in the figure. The reason of the error growing upon the $|k|$ increasing is that the probability amplitude flux at large energy $E$ is the rapidly oscillating function on $t$ (that is the product of $\exp(iEt)$ and
quite slowly changing functional of the wavefunction). The Fig.\ref{fig:h_convergence_t20}(a) demonstrates the error arising upon the using of Filon-Simpson's rule for the time integration of the probability amplitude flux that is the formula especially developed for the fast oscillating functions integration. It is seen that in contrast to the Simpson's rule applying results the error does not grow under $k$ increasing in the region of the large $|k|$ values. But the error grows a bit at small $|k|$.

Finally let us make some remarks upon the convergence of the PCG approach used for E-SURFF. During the TFECS using with $N=1024$ and preconditioner of the form \eqref{PreconFFT_ECS} with $N_l=10$ the average (over all the calculated spectrum points) iterations number was $N_{iter}\approx 79$ until the
discrepancy of the equation \eqref{linsys} was equal to $\varepsilon=10^{-12}$. In contrast, during the SECS using with the same grid and preconditioner parameters the average PCG iterations number was as small as $N_{iter}\approx 36$. The reason for such a situation is that $q(x)$ differs from 1 strongly in some points for TFECS, while it does not for the SECS. Thus in the case when small $E$ amplitudes are not needed it makes sense to prefer SECS more likely than TFECS. Whereas the calculation accuracy is not crucial one might utilize even the TFAP approach as it implies the PCG iterations number being as small as $N_{iter}\approx 20$ under using the preconditioner of the form \eqref{PreconFFT_ECS}.

\section{Conclusion}

In the present work we are developing the technique aimed to the combining of FFT and ECS approaches for the TDSE solution in the context of the important quantum mechanical and optical problems consideration. The technique consists in the TDSE solution by means of the split-operator method through the splitting of the kinetic energy operator into the coordinate-independent and coordinate-dependent parts. The coordinate-dependent part can be approximated by the finite-difference scheme.

In the present work the proposed technique efficiency is proved by means of the continuum amplitudes calculation through the t\&E-SURFF approach \cite{Serov2013}. The latter implies the solution of not only the TDSE, but of the SSE with the right-hand side as well. So we have also demonstrated that the solution of the SSE with the right-hand side by means of the conjugate gradient method may be performed with the help of the FFT-based preconditioner. 

Further, during the testing we investigated the ECS method efficiency with different contours in use, bearing in mind that the Fourier transform converges rapidly for a smooth function only. As a result we may conclude that the ECS approach can be accomplished by making use of the advanced TFECS contour proposed here as well. It provides the high degree of the unphysical reflection suppression for all energy values along with the small reflection from the approximated absorbing operator itself.

Finally, all the above mentioned actions taken resulted in the error decreasing such that the time integration error became distinguishable. So we also examined the efficiency of the numerical integration algorithm in the t-SURFF implementation. We can make a conclusion that the time integration of the probability amplitude flux is preferably to be performed by means of the Filon-Simpson's rule.

In summary, we have elaborated the highly efficient and fast method which can be helpful in solving a number of quantum mechanical and optical problems involving the evaluation of the amplitudes of ionization and dissociation of atoms and molecules by external fields, the laser beams propagation in the waveguides with leaking and so on. Thereupon, in order to apply the complex approach developed in the current article to the real problems, in the future we plan to upgrade it for a multidimensional case.

\begin{acknowledgments}
The authors would like to thank Prof. S. I. Vinitsky for fruitful discussions during the work on the paper. The authors acknowledge support of the work from the Russian Foundation for Basic Research (Grant No. 14-01-00520-a).
\end{acknowledgments}

\end{document}